# A density functional study of molecular oxygen adsorption and reaction barrier on Pu (100) surface

M. N. Huda and A. K. Ray*

*P.O. Box 19059, Department of Physics, The University of Texas at Arlington*
*Arlington, Texas 76019*

**Abstract.** Oxygen molecule adsorptions on a Pu(100) surface have been studied in detail, using the generalized gradient approximation to density functional theory. Dissociative adsorption with a layer by layer alternate spin arrangement of the plutonium layer is found to be energetically more favorable compared to molecular adsorption. *Hor*2 approach on a bridge site without spin polarization was found to the highest chemisorbed site with an energy of 8.787 eV among all the cases studied. The second highest chemisorption energy of 8.236 eV, is the spin-polarized *Hor2* or *Ver* approach at center site. Inclusion of spin polarization affects the chemisorption processes significantly, non-spin-polarized chemisorption energies being typically higher than the spin-polarized energies. We also find that the 5f electrons to be more localized in spin-polarized cases compared to the non-spin-polarized counterparts. The ionic part of O-Pu bonding plays a significant role, while the Pu 5f-O 2p hybridization was found to be rather week. Also, adsorptions of oxygen push the top of 5f band deeper away from the Fermi level, indicating further bonding by the 5f orbitals might be less probable. Except for the interstitial sites, the work functions increase due to adsorptions of oxygen.

## A. Introduction

Considerable theoretical efforts have been devoted in recent years to studying the electronic and geometric structures and related properties of surfaces to high accuracy. One of the many motivations for this burgeoning effort has been a desire to understand the detailed mechanisms that lead to surface corrosion in the presence of environmental gases; a problem that is not only scientifically and technologically challenging but also environmentally important. Such efforts are particularly important for systems like the actinides for which experimental work is relatively difficult to perform due to material problems and toxicity. As is known, the actinides are characterized by a gradual filling of the 5f-electron shell with the degree of localization increasing with the atomic number Z along the last series of the periodic table. The open shell of the 5f electrons determines the magnetic and solid-state properties of the actinide elements and their compounds and understanding the quantum mechanics of the 5f electrons is the defining

akr@uta.edu

issue in the physics and chemistry of the actinide elements. These elements are also characterized by the increasing prominence of relativistic effects. Studying them can, in fact, help us to understand the role of relativity throughout the periodic table. Narrower 5*f* bands near the Fermi level, compared to 4*d* and 5*d* bands in transition elements, are believed to be responsible for the exotic structure of actinides at ambient condition [1]. The 5f orbitals have properties intermediate between those of localized 4f and delocalized 3d orbitals and, as such, the actinides constitute the "missing link" between the d transition elements and the lanthanides [2]. Thus, a proper and accurate understanding of the actinides will help us understand the behavior of the lanthanides and transition metals as well.

Among the actinides, plutonium (Pu) is particularly interesting in two respects [3-6]. First, plutonium has, at least, six stable allotropes between room temperature and melting at atmospheric pressure, indicating that the valence electrons can hybridize into a number of complex bonding arrangements. Second, plutonium represents the boundary between the light actinides, Th to Pu, characterized by itinerant 5f electron behavior, and the heavy actinides, Am and beyond, characterized by localized 5f electron behavior. In fact, the high temperature fcc δ-phase of plutonium exhibits properties that are intermediate between the properties expected for the light and heavy actinides. These unusual aspects of the bonding in bulk plutonium are apt to be enhanced at a surface or in an ultra thin film of plutonium adsorbed on a substrate, due to the reduced atomic coordination of a surface atom and the narrow bandwidth of surface states. For this reason, plutonium surfaces and films and adsorptions on these may provide a valuable source of information about the bonding in plutonium.

This work has concentrated on square plutonium layers corresponding to the (100) surface of plutonium and adsorptions of molecular oxygen $O_2$, on such surfaces, using the formalism of modern density functional theory. Although the monoclinic α-phase of plutonium is more stable under ambient conditions, there are advantages to studying δ-like layers. First, a very small amount of impurities can stabilize δ-Pu at room temperature. For example, $Pu_{1-x}Ga_x$ has the fcc structure and physical properties of δ-Pu for $0.020 \leq x \leq 0.085$ [7]. Second, grazing-incidence photoemission studies combined with the calculations of Eriksson *et al.* [8] suggest the existence of a small-moment δ-like surface on α-Pu. Our work on plutonium monolayers has also indicated the possibility of such a surface [9]. Recently, high-purity ultrathin layers of plutonium deposited on Mg were studied by X-ray photoelectron (XPS) and high-resolution valence band (UPS) spectroscopy by Gouder *et al* [10]. They found that the degree of delocalization of the 5f states depends in a very dramatic way on the layer thickness and the itinerant character of the 5f states is gradually lost with reduced thickness, suggesting that the thinner films are δ-like.

Localised 5f states, which appear as a broad peak 1.6 eV below the Fermi level, were observed for one monolayer. At intermediate thickness, three narrow peaks appear close to the Fermi level and a comparative study of bulk α-Pu indicated a surface reorganization yielding more localized f-electrons at thermodynamic equilibrium. Finally, it may be possible to study 5f localization in plutonium through adsorptions on carefully selected substrates for which the adsorbed layers are more likely to be δ-like than α-like.

The anomalous properties of δ-Pu have triggered extensive studies on its electronic structures and ground state properties over the years. Different levels and types of theories have been proposed and used to deal with this strongly correlated system. Standard density functional theory (DFT), which works well for the lighter actinides, was found to be inadequate to for the description of some of the ground state properties of δ-Pu [11]. For example, DFT in the local density approximation (LDA) for the electron exchange and correlation effects underestimates the equilibrium volume up to 30% and predicts an approximately four times too large bulk modulus [12-13]. The electronic structure is, in fact, incompatible with photoemission spectra. On the other hand, theories beyond LDA, such as, the self-interaction-corrected (SIC) LDA studied by Petit *et al.* [14] predicted a 30% too large equilibrium volume. Penicaud [15] performed total energy calculations in the local density approximation using fully relativistic muffin-tin orbital band structure method. For δ-Pu, the $5f_{5/2}$ electrons were uncoupled from the s, p and d electrons to reproduce the experimental value of the equilibrium atomic volume. Also an adjustable parameter was introduced to get a better theoretical representaion of δ-Pu. Using 'mixed-level' model, where the energies were calculated at both localized and delocalized 5f configurations, Eriksson *et al.* [16] reproduced reseasonable equilibrium volumes of U, Pu and Am. There have been also attempts to use the LDA+U method, where U is the adjustable Hubbard parameter, to describe the electron correlation within the dynamical mean field theory (DMFT) [17]. The experimental equilibrium δ-Pu volume was reproduced, with U equal to 4 eV.

As is known, the existence of magnetic moments in bulk δ-Pu is also a subject of great controversy and significant discrepancies exist between various experimental and theoretical results. To this end, we comment on a few representative works in the literature, partly to mention explicitly some of the controversies. Susceptibility and resistivity data for δ-Pu were published by Meot-Reymond and Fournier [18], which indicated the existence of small magnetic moments screened at low temperatures. This screening was attributed to the Kondo effect. Recent experiments by Curro and Morales [7] of 1.7 percent Ga-doped δ-Pu conducted at temeperatures lower than the proposed Kondo Temperature of 200-300 K showed little evidence for local magnetic moments at the Pu sites. Though there is no direct evidence for magnetic moment, spin-polarized DFT, specifically the generalized-gradient-approximation (GGA) to DFT, has been used by theoreticians, in particular, to predict the magnetic ordering and the ground state properties of δ-Pu. This is partly due to the fact that spin-polarized DFT calculations do predict better agreement with photoemission data. Basically, inclusion of the spin polarization enhance the electron localization to plutonium atoms, which is needed for the description of δ-Pu. Niklasson *et al*. [19] have presented a first- principles disordered local moment (DLM) picture within the local-spin-density and coherent potential approximations (LSDA+CPA) to model some of the main characteristics of the energetics of the actinides, including δ-Pu. The authors also descibed the failures of the local density approximation (LDA) to describe 5f localization in the heavy actinides, including elemental Pu. The DLM density of states was found to compare well with photoemission on δ-Pu, in contrast to that obtained from LDA or the magnetically ordered AFM configuration. On the other hand, Wang and Sun [20], using the full-potential linearized augmented-plane-wave (FP-LAPW) method within the spin-polarized generalized gradient approximation (SP-GGA) to density functional theory, without spin-orbit coupling, found that that the antiferromagnetic-state lattice constant and bulk modulus agreed better with experimental values than the nonmagnetic values of δ-Pu. Using the fully relativistic linear combinations of Gaussian-type orbitals-fitting function (LCGTO- FF) method within GGA, Boettger [21] found that, at zero pressure, the AFM (001) state was bound relative to the non-magnetic state by about 40 mRy per atom. The lattice constant for the AFM (001) state also agreed better with the experimental lattice constant as compared to the nonmagnetic lattice constant. However, the predicted bulk modulus was significantly larger than the experimental value. Söderlind *et al.* [22], employing the all electron, full-potential-linear-muffin-tin-orbitals (FLMTO) method, predicted a mechanical instability of antiferromagnetic δ-Pu, and proposed that δ-Pu is a 'disordered magnet'. In a more recent study on 5f localization, Söderlind *et al.* showed that 5f-band fractional occupation at 3.7 (68% atoms with itinerant 5f electrons) can predict well the atomic volume and bulk modulus without referring to the magnetic ordering. Wills *et al.* [23] have claimed that there is, in fact, no evidence of magnetic moments in the bulk δ- phase, either ordered or disordered. Using the full-potential-linearized-augmented-plane-wave (FP-LAPW) method, Wu and Ray [24] have calculated the equilibrium atomic volume, 178.3 a.u.$^3$ and bulk modulus 24.9 GPa of ferromagnetic bulk δ-Pu at the fully relativistic level of theory, in good agreement with the experimental values of 168.2 a.u.$^3$ and 25 GPa (593 K), respectively.

As mentiond before, oxidation of plutonium is important for many reasons, including scientific, technological, and environmental such as the long-term storage of plutonium. Experimental data [25] indicates that when plutonium surface is exposed to molecular oxygen, oxygen is readily adsorbed by the metal surface. The oxygen molecule then dissociates into atomic oxygen, and combines with plutonium to form a layer of oxide. Oxidation continues and the oxygen diffuses through the oxide layer reacting with more plutonium and producing more oxide at the oxide/metal interface, eventually reaching a steady state thickness. Almeida *et al.* [26] studied the adsorption of $O_2$, $CO_2$, CO and $C_2H_4$ on plutonium metal at 77 and 296 K by UPS and XPS. For $O_2$ adsorption, they showed that initially $Pu_2O_3$ is formed, which is then followed by an oxidation to $PuO_2$. Using the film-linearized-muffin-tin-orbitals (FLMTO) method, Eriksson *et al.* [8] have studied the electronic structure of hydrogen and oxygen chemisorbed on plutonium. The slab geometry was chosen to have the $CaF_2$ structure and the chemisorbed atoms were assumed to have fourfold-bridging positions at the surface. They found the surface behavior in $PuO_2$ to be rather different compared to the surface behavior in pure metallic plutonium. For metallic plutonium, the 5f electrons are valence electrons and show only a small covalent like bonding contribution associated with small 5f to non-5f band hybridization. For the oxide, the Pu 5f electrons were well localized and treated as core electrons. Thus, the plutonium valence behavior is dominated by the 6d electrons, giving rise to significant hybridization with ligand valence electrons and significant covalency. Huda and Ray [27] have recently studied *atomic* oxygen adsorption on $\delta$-Pu (100) and (111) surfaces at both non-spin-polarized and spin-polarized levels using the generalized gradient approximation of density functional theory (GGA-DFT) with Perdew and Wang (PW) functionals [28-29] . The center position of the (100) surface was found to be the most favorable site with chemisorption energies of 7.386eV and 7.080eV at the two levels of theory. For the (111) surface non-spin-polarized calculations, the center position was also the preferred site with a chemisorption energy of 7.070eV, but for spin-polarized calculations the bridge and the center sites are found to be basically degenerate, the difference in chemisorption energies being only 0.021eV. In our previous hybrid density functional cluster study of the bulk and surface electronic structures of PuO [30], a large overlap between the Pu 5f bands and O 2p bands and a significant covalent nature in the chemical bonding were found. The highest occupied molecular orbital – lowest unoccupied molecular orbital (HOMO-LUMO) gaps and the density of states of the clusters supported the idea that PuO is a semiconductor. In follow-up studies of $PuO_2$ (110) surface and water adsorption on this surface, we have shown that the adsorption is dissociative and oxygen interaction is relatively strong. In a recent study using the self-interaction corrected local spin density method, Petit *et al.* [31] reported the electronic structure of $PuO_{2\pm x}$. They found that in the stoichiometric $PuO_2$ compound, Pu occurs in the Pu (IV) oxidation state, corresponding to a localized $f^4$ shell. If oxygen is introduced onto the octahedral interstitial site, the nearby Pu atoms turn into Pu (V) ($f^3$) by transferring electron to the oxygen. We also wish to mention that *no detailed information exists in the literature about the magnetic state of the molecule-adsorbed surface of plutonium and our present study including spin polarization on molecular oxygen adsorptions on plutonium surfaces is a first step towards an understanding of molecular chemisorption on Pu surfaces and the influence of surface magnetism*. We also note that, as the films get thicker, the complexity of magnetic ordering, if existent, increases and such calculations are quite challenging computationally. Nevertheless, to study the effects of spin polarization on the chemisorption process, our studies have been performed at both the spin-polarized and at the non-spin-polarized levels.

**B. Computational details**

As in our previous works [27,30], all computations reported here have been performed at both the spin unrestricted and the spin restricted generalized gradient approximation (GGA) level of density functional theory (DFT) [28-29], using the suite of programs DMol3 [32]. This code does not yet allow fully relativistic computations and, as such, we have used the scalar-relativistic approach. In this approach, the effect of spin-orbit coupling is omitted primarily for computational reasons, but all other relativistic kinematic effects, such as mass-velocity, Darwin, and higher order terms are retained. It has been shown [32] that this approach models actinide bond lengths fairly well. We certainly do not expect that the inclusion of the effects of spin-orbit coupling, though desirable, will alter the primary qualitative and quantitative conclusions of this paper, particularly since we are interested in chemisorption energies defined as the difference in total energies and it is expected that the shift in total energies in Pu and Pu+$O_2$ system due to inclusions of spin-orbit coupling is expected to basically cancel each other. We also note that Landa *et al.* [22] and Kollar *et al.* [33] have observed that inclusions of spin-orbit coupling are not essential for the quantitative behavior of δ - Pu. Hay and Martin [34] found that one could adequately describe the electronic and geometric properties of actinide complexes without treating spin-orbit effects explicitly. Similar conclusions have been reached by us in our study of water adsorption [30] and of molecular $PuO_2$ and $PuN_2$ [35] and by Ismail *et al*. [36] in their study of uranyl and plutonyl ions. We also note that scalar-relativistic hybrid density functional theory has been used by Kudin *et al*. [37] to describe the insulating gap of $UO_2$, yielding a correct antiferromagnetic insulator.

In DMol3, the physical wave function is expanded in an accurate numerical basis set, and fast convergent three-dimensional integration is used to calculate the matrix elements occurring in the Ritz variational method. For the oxygen atom, a double numerical basis set with polarization functions (DNP) and real space cut-off of 5.0 Å was used. The sizes of these DNP basis set are comparable to the 6-31G** basis of Hehre *et al.* [38]. However, they are believed to be much more accurate than a Gaussian basis set of the same size [32]. For Pu, the outer sixteen electrons ($6s^2\ 6p^6\ 5f^6\ 7s^2$) are treated as valence electrons and the remaining seventy-eight electrons are treated as core. A hardness conserving semi-local pseudopotential, called density functional semi-core pseudo-potential (DSPP), has been used. These norm-conserving pseudo-potentials are generated by fitting all-electron relativistic DFT results and have a non-local contribution for each channel up to $l = 2$, as well as a non-local contribution to account for higher channels. To simulate periodic boundary conditions, a vacuum layer of 30 Å was added to the unit cell of the layers. The k-point sampling was done by the use of the Monkhorst-Pack scheme [39]. The maximum number of numerical integration mesh points available in DMol3 has been chosen for our computations, and the threshold of density matrix convergence is set to $10^{-6}$. All computations have been performed on a Compaq ES40 alpha multi-processor supercomputer at the University of Texas at Arlington.

C. Results and discussions

As in our earlier study of $O_2$ adsorption on uranium surfaces [40], to study $O_2$ adsorption on the plutonium (100) surface, we have modeled the surface with three layers of fcc plutonium at the experimental lattice constant. One of the reasons for choosing the experimental lattice constant comes from the fact that as mentioned above, controversies abound in direct applications of DFT to computations of atomic volume and bulk modulus of Pu and another reason comes from our wish to simulate the experimental chemisorption process as much as possible. The choice of three layers is believed to be quite adequate considering that the oxygen molecule is not expected to interact with atoms beyond the first three layers. This was found to be the case in our studies of atomic oxygen and hydrogen adsorptions on the plutonium surface [27, 41]. Recently, Ray and Boettger [42] showed in a study of quantum size effects of δ-plutonium surface that the surface energies converge within the first three layers. The unit cell for our study here is chosen to contain four plutonium atoms per layer to provide a very accurate representation of the molecular adsorption process. Thus our three-layer model of the surface contains twelve plutonium atoms. For spin-polarized calculations, spin arrangements of the plutonium atoms of the bare (100) surface was optimized and the arrangements with lower total energy was used for oxygen adsorption calculations. The $O_2$ molecule, one per unit cell, was allowed to approach this plutonium surface along three different symmetrical sites: i) directly on top of a plutonium atom (*top* site); ii) on the middle of two nearest neighbor plutonium atoms (*bridge* site); iii) in the center of the smallest unit structures of the surfaces (*center* site). As the smallest structure of (100) δ-like plutonium surface is a square, these three sites are the only symmetrically distinguishable sites. In addition to this, we have also considered some positions inside the Pu three layers slab (*interstitial* positions). For each of these positions, we consider several approaches for chemisorptions. They are: *a*) $O_2$ molecule approach is vertical to the surface (*Ver* approach), *b*) $O_2$ molecule approach is parallel to the surface and parallel to the square lattice vectors (*Hor1* approach), and *c*) $O_2$ molecule approach is parallel to the surface and at an angle $45^o$ with the square lattice vectors, (*Hor2* approach). It is obvious that for both of the horizontal approaches the atoms of the oxygen molecule $O_2$ are at the same distance from the plutonium surface, whereas for the vertical approach one oxygen atom is closer to the surface than the other. For geometry optimizations, the distances of the oxygen atoms from the surface and the distance between the oxygen atoms ($r_O$) were simultaneously optimized. The chemisorption energies were then calculated from:

$$E_c = E(\text{Pu-layers}) + E(O_2) - E(\text{Pu-layers} + O_2) \quad (1)$$

A positive chemisorption energy thus implies the probability of chemisorption. For the non-spin-polarized case, both E (Pu-layers) and E (Pu-layers + $O_2$) were calculated without spin polarization, while for spin polarized chemisorption energies both of these energies are spin polarized. $E(O_2)$ is the energy of the oxygen molecule in the ground state. The chemisorption energies, and the corresponding distances are given in table 1. The distances $r_d$ given in the tables are measured as the distance from the plutonium surface to the oxygen atoms if both the oxygen atoms are at same height or to the nearer oxygen atom if one of them is closer to the surface than the other.

It is well known that oxygen molecule adsorption on metal surface is strong, and the adsorption is usually dissociative. Thus, one purpose of this work is to investigate the probability of dissociative adsorption compared to molecular adsorption. We start by describing the chemisorption processes of $O_2$ at the different sites on plutonium surfaces and discuss the spin-magnetic properties of these adsorptions followed by a study of the reaction barrier for the dissociation of $O_2$ on plutonium surfaces. The effects of oxygen adsorptions on the nature of plutonium 5f orbitals are also described below.

We first discuss the top sites without spin polarization. Figure 1 has the optimized $O_2$ chemisorbed geometries on plutonium surface for top positions. It was mentioned earlier that there are three different approaches for each site. For the two horizontal approaches the chemisorption parameters are almost the same, namely the distances ($r_d$) from the plutonium surface to the $O_2$ are 2.01 Å and 1.99 Å, respectively and the O-O bond lengths are stretched up to 1.49 Å and 1.51 Å from the experimental bond length of 1.21 Å; but the chemisorption energies are slightly different, 3.128 eV and 3.414 eV for *Hor1* and *Hor2*, respectively. In these cases, both the oxygen atoms coordinated with the plutonium atom below them. It is noted that, as the other parameters are almost the same, a rotation of 45º will transform the *hor1* approach to the *hor2* approach with an increment of chemisorption energy of 0.286 eV. For the *Ver* approach the distance of the lower oxygen atom to the plutonium surface is 2.03 Å, with the lowest chemisorption energy of 2.099 eV. For all these three approaches, the $O_2$ adsorption is molecular, the maximum increase in O-O bond length being 0.30 Å for *Hor2* approach from the experimental bond length of $O_2$, while for *Ver* approach the O-O bond length increases by only 0.10 Å. Inclusion of spin polarization bring the $O_2$ 0.16 Å nearer to the plutonium surface for *Hor1* approach, but the other over all geometric features for the top site for all the three approaches did not change significantly, though the spin polarized adsorptions have lower chemisorption energies.

For the bridge sites, the chemisorptions of $O_2$ along the vertical approach behaved differently for non-spin-polarized and spin-polarized cases. For the non-spin-polarized case, $O_2$ remained as a molecule, while for spin-polarized case the oxygen molecule dissociated and the final adsorption sites resembled to the top site at *Hor1* approach as if the oxygen molecule was dissociated (figure 2). The O-O optimized distance for this approach is 3.03 Å, and the chemisorption energy has a rather high value of 7.166 eV. For the *Hor1* approach at bridge sites, $O_2$ completely dissociates for the non-spin-polarized case and each oxygen atom sits on the two top layer plutonium atoms, with a chemisorption energy of 6.647 eV. The inclusion of spin polarization could not break up the oxygen molecule ($r_O$ is 1.49 Å), and hence give lower chemisorption energy with a lower distance from the plutonium layer than the corresponding non-spin-polarized case. For the *Hor2* approach the oxygen molecule dissociate and the spin polarization does not have any considerable effect on chemisorption geometry. However the spin-polarized adsorption energy is 0.854 eV lower than its non-spin-polarized counterpart. For *Hor2* approach, after dissociation, each oxygen atom sits on the two adjacent center positions. The non-spin-polarized chemisorption energy of *Hor2* is the highest among the all other chemisorption sites and approaches considered here. In general, chemisorption at the bridge site is considerably stronger than at the top site. This results from fact that oxygen atoms are relatively much closer to the plutonium surface in bridge sites compared to the top sites. However, we note that geometrically the difference of spin polarized *Hor1* approaches between these two sites, where the adsorption is molecular, is only a small translational shift with a difference of chemisorption energy of 1.421 eV. In top position both oxygen atoms are coordinated with the same plutonium atom, whereas in bridge position oxygen atoms are coordinated with different plutonium atoms.

For the center site at *Ver* approach, like bridge sites, in spin-polarized case $O_2$ dissociated completely with O-O distance of 4.29 Å, while for the non-spin polarized case $r_O$ is 1.47 Å. The final optimized spin polarized *Ver* approach is identical, in geometry and in chemisorption energies, with the spin polarized optimized *Hor2* approach of center site (figures 3 (a) and (c)). The chemisorption energy of these sites is 8.236 eV, which is the second highest among the chemisorption configurations studied here. However, the non-spin-polarized *Ver* approach, where the $O_2$ did not dissociate, the chemisorption energy is as low as 2.939 eV comparable to the adsorption energies of the top sites. The non-spin-polarized *Hor2* approach, with O-O bond length of 2.72 Å, has the chemisorption energy of 7.216 eV. For *Hor1* approach, where after dissociation the oxygen atoms sit almost in two neighboring bridge sites (figure 3(b)), the chemisorption energies are 8.100 eV and 7.171 eV for without and with spin-polarization, respectively, and the chemisorbed distances and O-O distances are identical in both the two cases.

The adsorption distances, $r_d$, are usually lower for the center site, and in general the chemisorption energies are higher than the other two sites, except for the *Hor2* approach in center and bridge site, where the bridge site has lower $r_d$ than the center site, but still spin polarized center site has higher chemisorption energies. We know from our atomic oxygen adsorption study on plutonium surface [27] that the center is the most favorable atomic adsorption site. In this present study, in spin-polarized bridge site at *Hor2* approach, after dissociation of $O_2$, each oxygen atom goes into adjacent center positions. However, in the case of center site at *Hor2* approach each oxygen atom goes into diametrically opposite center positions and hence has higher O-O distances. As always oxygen atoms gain significant amount of charges (in these cases ~0.64$e$) from the plutonium surface atoms in the chemisorption processes, and the higher distances between oxygen atoms lower the coulomb repulsion forces between them, which explains the higher chemisorption energies in spin polarized *Hor2* approach of center site than that of bridge site. Similar arguments can be made for the non-spin-polarized bridge and center sites. However, the same argument does not hold if we compare between the *Hor2* approaches of non-spin–polarized and spin-polarized bridge site with that of spin polarized center site. This is because of the inclusion of spin magnetic effect. It is noted that the final optimized position of oxygen atoms on plutonium surface for spin-polarized *Ver* approach and spin-polarized *Hor2* approach for center site are almost the same, and similar to the *Hor2* approach of top site if the $O_2$ was dissociated.

Different interstitial positions in the above discussed symmetrical sites were also studied, and many of the sites and approaches yield negative chemisorption energies, *i.e.*, the oxygen molecule cannot be adsorbed for those sites. However, our calculations did yield positive chemisorption energy when the $O_2$ molecule dissociates and sits near the middle of the three layers for both with and without spin polarization. The corresponding $r_d$ and $r_O$ along with chemisorption energies are provided in table 1. However, the chemisorption energy of 4.33eV (for non-spin-polarized case) is almost half of the chemisorption energy for the most stable site above the surface, which is 8.787 eV for the non-spin-polarized *Hor2* approach of bridge site. Similar comments apply to the spin-polarized case, where the chemisorption energy is less than half than the energy for the center site. We can infer that, at the initial stage of oxidation on plutonium surfaces, dissociated oxygen atoms form a layer on the surface, before diffusion into the bulk to form plutonium oxide.

From the above discussions and table 1, it is clear that vertical approaches where the $O_2$ adsorption is molecular have significantly lower chemisorption energies compared to the other cases where $O_2$ dissociates. Basically in molecular adsorption at vertical approach, one oxygen atom that is closer to the plutonium surface is coordinated with plutonium surface atoms, while the other one is only bonded with oxygen atom; whereas in other cases both oxygen atoms are bonded with the surface plutonium atoms. This explains the much lower chemisorption energies of the vertical approaches despite the fact that oxygen atoms are sometimes much closer to the surface, *e.g.*, at center sites, where, for non-spin-polarized case, the first oxygen atom is at 0.87 Å from the plutonium surface and the second one is at 2.34 Å, *i.e.*, the second atom is at 3.17 Å from the nearest plutonium atom. However, electronic charge on the lower atom (−0.37$e$) is slightly smaller than the higher atom (−0.41$e$). This feature is also true for other vertical approaches where the adsorption is molecular. For example in top site of non-spin polarized case at *Ver* approach, the difference in charges on oxygen atoms is the largest. The lower atom has −0.16$e$ and the higher one have the charge of −0.25$e$. Here the distance of the nearest plutonium atom from the higher oxygen atom is 3.33 Å. The difference in charges is minimum for non-spin-polarized bridge site where the charges on the lower and higher oxygen atoms are −0.31$e$ and −0.32$e$, respectively, with the nearest Pu-O distance from the higher atom is 3.33 Å. The plutonium surface basically interacts with the first oxygen atom, while the coordination with the second atom to the surface may be screened by the oxygen atom nearer to the surface.

Table 2 lists the Mulliken charge distributions [43] of the bare and the most stable chemisorption sites for both the non-spin-polarized and the spin-polarized cases, namely the *Hor2* approaches of the bridge and the center sites, respectively. The overall charge distribution patterns for the both chemisorbed sites are almost same. Before the adsorption of oxygen, plutonium atoms on the first layer, as well as in the third layer, were slightly negatively charged, while the atoms on second layer were positively charged. This particular symmetric pattern of charge distribution reflects the symmetry of the unit cell chosen for the calculations. After the adsorption of oxygen molecule, *e.g.*, in spin polarized case, the oxygen atoms acquire negative charges, $-0.646e$ each, primarily from the plutonium atoms on the first layer, yielding a positively charged top layers. As a result, the total charge transfer to $O_2$ is $-1.292e$ from the plutonium slab (for non-spin polarized case the transferred charge is $-1.138e$), and hence there exists a strong ionic part in the Pu-O bonding, along with other contributions. The plutonium atoms in the second layer which are directly below the oxygen atoms are more positive than the surrounding plutonium atoms. Also, we find that in the oxygen adsorbed plutonium layers for non-spin-polarized case, the second and third layers charge distributions are slightly modified compared to the bare cases. For spin-polarized case, the spin arrangements on the surface affect the charge distribution, though the overall pattern remains the same. However, from the fact that the second and third layers of both non-spin-polarized and spin-polarized oxygen adsorbed surfaces remain positively and negatively charged as the bare plutonium layers, we can infer that the effect of oxygen adsorption is minimal beyond the third layer.

In table 1, the magnetic moments of the oxygen adsorbed plutonium layers for different adsorption configurations are tabulated. It is well known that due to the reduced dimensions and narrow electronic states, even the surfaces of paramagnetic bulk may have magnetic moments. It has been mentioned earlier that the magnetism in plutonium has been a source of great controversies in recent years. In this study, the average magnetic moment of bare plutonium layers is found to be 2.090 $\mu_B$ per atom with layer by layer alternating spin. The first and third layers have up spins, while the second layer has down spins. It was found that the adsorption of oxygen molecule did not change the spin moment significantly, giving an average value of 1.658 or 1.681 $\mu_B$ per atom for the most preferred chemisorption sites. Moments on the adsorbed oxygen atoms are very small. The magnetic moments as shown in table 1 have about the same magnitude, and lack any specific orderings. Table 2 has the distribution of spins for bare plutonium layers, which shows basically an anti-ferromagnetic behavior, in agreement with some theoretical calculations for bulk Pu [20, 21]. Other spin arrangements yield higher total energy than the anti-ferromagnetic one. Table 2 also shows the spin distribution of the center site at *Ver* or *Hor2* approach, which shows an almost layer by layer alternate spin arrangement like the bare plutonium surface, which again might be a precursor of anti-ferromagnetic behavior. The same is true for all the other sites of spin polarized cases. In our earlier study of $O_2$ adsorption on uranium layer [40], we found that spin magnetic moment did not play a significant role on chemisorption processes. In contrast here, spin polarization does have a significant effect on chemisorption processes, consistent with our previous studies of oxygen and hydrogen atom adsorption on plutonium surfaces [27, 41].

A study of the energy levels of the plutonium layers before oxygen adsorption indicates that while the plutonium 6s and 6p electrons are localized, a fraction of 5f electrons appear to be delocalized. From the band energetics of the bare and oxygen adsorbed plutonium layers, we also found that the change in band gaps due to the inclusion of spin polarization is significant. For example the energy differences for the top of 5f bands of plutonium (100) surface and the Fermi energy without spin polarization is 0.273eV, to be compared with the spin-polarized value of 0.514eV. Also the occupation number for the top energy level of 5f band is 0.774 and 0.958 without and with spin polarization, respectively. These and the magnetic moment considerations, as described above, indicate higher localization of 5f electrons when spin polarization is included. For the oxygen adsorbed layers, for the sake of brevity, we discuss only the most favored chemisorption configurations, *Hor2* approaches of bridge site and center site for non-spin-polarized and spin-polarized cases, respectively. The energy differences for the top of 5f bands and the Fermi energy for these sites are 0.294 eV and 0.539 eV, respectively. It can be inferred from these energies that the adsorptions of oxygen push the 5f band a little deeper from the Fermi energy, approximately by 0.02 eV. This may imply that the adsorption of oxygen inhibits the plutonium 5f orbitals to participate in further bonding. So the conduction band is basically composed of 6d and 7s electrons. This may also be a precursor of semiconducting behavior of $PuO_2$. In figures 4 and 5 we have plotted 5f-DOS for plutonium (100) bare surfaces and the most favorable oxygen chemisorbed surfaces at both non-spin-polarized and spin-polarized levels, respectively. A Gaussian broadening procedure has been employed here to compute the DOS [26]. A Gaussian $\exp(-\alpha x^2)$ is assigned to each energy eigenvalue with $\alpha = 1000$, such that the width at the half height is 0.05 eV. From the DOS it is clear that the hybridization between the plutonium 5f orbitals and the oxygen 2p orbitals is rather weak and the bonding between the oxygen and the plutonium surface is mainly ionic. Also the overall pattern of the DOS is affected by oxygen adsorption. In the oxygen adsorbed non-spin polarized DOS plot, there is an energy gap between approximately −4.5 eV and −3.5 eV between the hybridized 5f-2p orbitals and the remaining 5f orbitals. For the spin polarized case, the gap is smaller and between approximately −3.7 eV and −3.0 eV. Change in Fermi energy due to the adsorption of oxygen is 0.233 eV and 0.424 eV without and with spin polarization, respectively, for the corresponding most favorable chemisorption sites. Thus, the work function increases due to the oxygen adsorption. This holds true for all the chemisorbed sites above the surfaces. Only for the interstitial positions, the presences of oxygen lower the work functions.

From the above discussions of the chemisorption processes, it is evident that the dissociative adsorption is favored over molecular adsorption. We also studied reaction barrier in the dissociation processes by constraint minimization of energy along a chosen reaction coordinate. Only the most favorable chemisorption sites are considered at both the non-spin-polarized and the spin-polarized levels. As for the reaction coordinate, we have chosen the O-O distances, starting at the experimental bond length. For reaction barrier calculations, $O_2$ was placed above the plutonium surface with the O-O distance kept fixed at a given value, and geometry optimization was done to yield the total energy of the system. Figure 6 shows the optimized energy curves with respect to the different O-O distances. The non-spin-polarized curve (*Ho2* approach at bridge site) shows a possibility of molecular adsorption of $O_2$ at experimental bond length with a chemisorption energy of 6.393 eV, compared to the dissociative adsorption of 8.787 eV, as reported above. However, the curve has a peak at O-O distance of 1.3 Å, with adsorption energy of 3.448 eV. From the molecular adsorption to the complete dissociative adsorption, there exists a small energy hill of 0.108eV. As all along the curve oxygen atoms are bound to the surface, this small energy hill does not pose any significant barrier to the $O_2$ dissociative process on plutonium surface. Among the spin-polarized cases both the *Ver* and the *Hor2* approaches on the center site has the same highest chemisorption energies. However, the initial dissociation of $O_2$ at *Hor2* approach occurs at a higher coordination where both the oxygen atoms are at the same proximity to the plutonium atoms on the surface compared to the *Ver* approach where one oxygen atom is nearer to the surface than the other. Thus, we only considered the dissociation at the *Hor2* approach. The spin-polarized curve shows no energy hill like its non-spin-polarized counterpart. However, it does indicate a possible molecular adsorption at O-O bond length of 1.5 Å. The adsorption energy here is 3.653 eV, much lower compared to the complete dissociative oxygen chemisorption energy of 8.236 eV.

**D. Conclusions**

In conclusion, our study of oxygen molecule adsorptions on Pu(100) surface, using the generalized gradient approximation to density functional theory, shows that the adsorption is dissociative. A layer by layer alternate spin arrangement of the plutonium layer is energetically most favorable and adsorption of oxygen does not change this feature. *Hor2* approach on bridge site without spin polarization with a chemisorption energy of 8.787 eV was found to be the highest chemisorbed site among all the cases studied here. The second highest chemisorption energy of 8.236 eV, is the spin-polarized *Hor2* or *Ver* approach at center site. Inclusion of spin polarization affects the chemisorption processes significantly, non-spin-polarized chemisorption energies being typically higher than the spin-polarized energies. We also find that 5f electrons are more localized in spin polarized case, than the non-spin polarized counterparts. The ionic part of O-Pu bonding plays a significant role, while the Pu 5f-O 2p hybridization was found to be rather week. Also adsorption of oxygen push the top of 5f band a little deeper from the Fermi level, indicating further bonding by the 5f orbitals might be less probable. Except for the interstitial sites, the work functions increase due to adsorptions of oxygen.

**Acknowledgments**

This work is supported by the Chemical Sciences, Geosciences and Biosciences Division, Office of Basic Energy Sciences, Office of Science, U. S. Department of Energy (Grant No. DE-FG02-03ER15409) and the Welch Foundation, Houston, Texas (Grant No. Y-1525).

Table 1. Chemisorption energies of $O_2$ adsorption on Pu (100) surface; $r_d$ and $r_O$ are the distances of oxygen atom from the Pu surface and the O-O distances, respectively.

| Sites | Approach | $r_d$ (in Å) | $r_O$ (in Å) | Chemisorption Energy (in eV) | Magnetic Moment per atom (in $\mu_B$) |
|---|---|---|---|---|---|
| | | | No Spin Polarization | | |
| Top | Ver | 2.03 | 1.31 | 2.099 | |
| | Hor1 | 2.01 | 1.49 | 3.128 | |
| | Hor2 | 1.99 | 1.51 | 3.414 | |
| Bridge | Ver | 1.61 | 1.36 | 2.285 | |
| | Hor1 | 1.81 | 3.03 | 6.647 | |
| | Hor2 | 0.85 | 3.03 | 8.787 | |
| Center | Ver | 0.87 | 1.47 | 2.939 | |
| | Hor1 | 1.38 | 3.03 | 8.100 | |
| | Hor2 | 1.67 | 2.72 | 7.216 | |
| Inter* | | 2.14 | 3.03 | 4.334 | |
| | | | With Spin Polarization | | |
| Top | Ver | 2.06 | 1.30 | 1.871 | 1.675 |
| | Hor1 | 1.85 | 1.47 | 2.926 | 1.682 |
| | Hor2 | 2.01 | 1.47 | 3.127 | 1.676 |
| Bridge | Ver | 1.42 | 3.03 | 7.166 | 1.585 |
| | Hor1 | 1.72 | 1.49 | 4.347 | 1.674 |
| | Hor2 | 0.87 | 3.04 | 7.933 | 1.681 |
| Center | Ver | 1.00 | 4.29 | 8.236 | 1.681 |
| | Hor1 | 1.39 | 3.03 | 7.171 | 1.581 |
| | Hor2 | 1.00 | 4.29 | 8.236 | 1.658 |
| Inter* | | 2.14 | 3.03 | 3.634 | 1.891 |

*Bridge, *hor1*

Table 2. Charge and spin distributions of bare plutonium layers and the most favorable chemisorption configurations for non-spin polarized (NSP, *Hor2* of bridge site) and spin polarized (SP, *Hor2* of center site) cases.

|  | Plutonium layers | | | Plutonium + Oxygen layers | | |
|---|---|---|---|---|---|---|
|  | NSP | SP | | NSP | SP | |
|  | Charge | Charge | Spin | Charge | Charge | Spin |
| O-atom | X | X | X | -0.569 | -0.646 | -0.126 |
|  | X | X | X | -0.569 | -0.646 | -0.126 |
| layer1 | -0.108 | -0.086 | 5.728 | 0.211 | 0.279 | 5.353 |
|  | -0.108 | -0.086 | 5.728 | 0.211 | 0.280 | 5.343 |
|  | -0.108 | -0.086 | 5.728 | 0.211 | 0.279 | 5.348 |
|  | -0.108 | -0.086 | 5.728 | 0.211 | 0.279 | 5.348 |
| layer2 | 0.216 | 0.172 | -5.19 | 0.291 | 0.191 | -5.267 |
|  | 0.216 | 0.172 | -5.19 | 0.144 | 0.191 | -5.267 |
|  | 0.216 | 0.172 | -5.19 | 0.290 | 0.141 | -5.169 |
|  | 0.215 | 0.172 | -5.19 | 0.144 | 0.141 | -5.169 |
| layer3 | -0.108 | -0.086 | 5.728 | -0.144 | -0.122 | 5.737 |
|  | -0.108 | -0.086 | 5.728 | -0.144 | -0.122 | 5.737 |
|  | -0.108 | -0.086 | 5.728 | -0.144 | -0.122 | 5.737 |
|  | -0.108 | -0.086 | 5.728 | -0.144 | -0.122 | 5.737 |

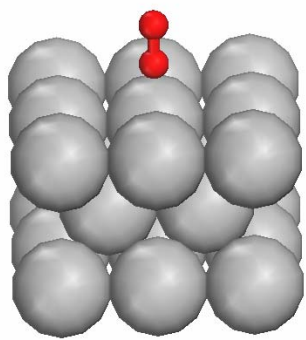 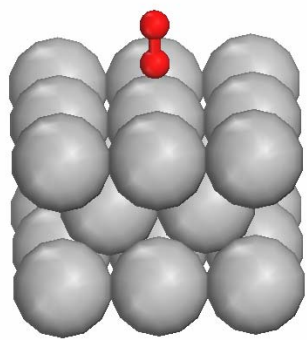

NSP  SP

(a) Ver approach.

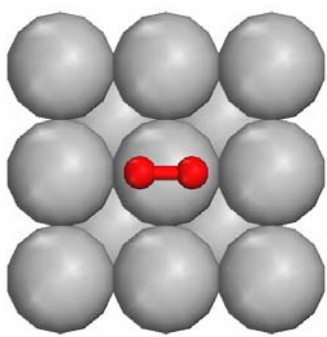 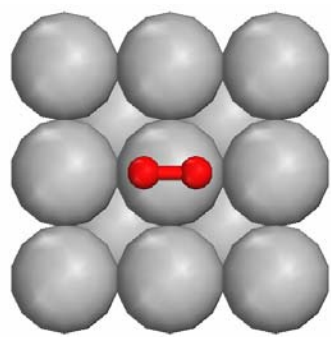

NSP  SP

(b) Hor1 approach.

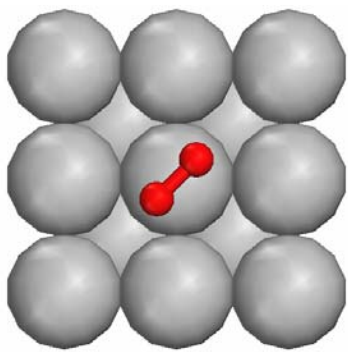 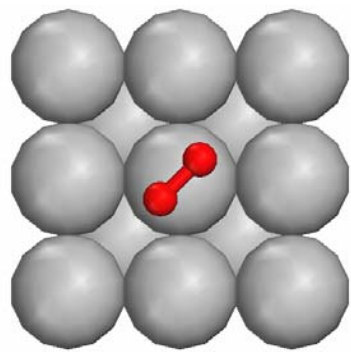

NSP  SP

(c) Hor2 approach.

Figure 1. Optimized configuration of $O_2$ adsorption on top site at different approaches. NSP and SP refer to non-spin and spin polarization, respectively.

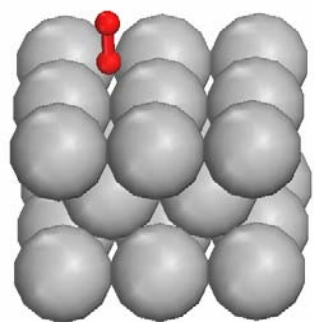
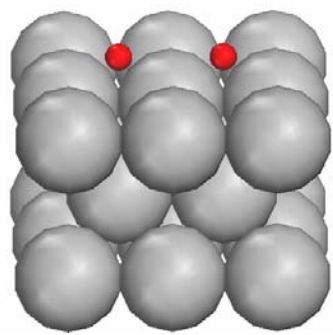

NSP  SP

(a) Ver approach.

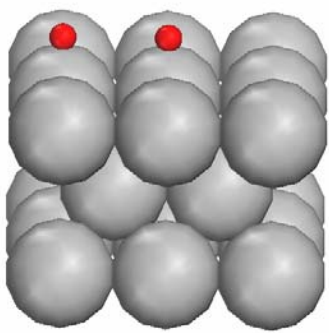
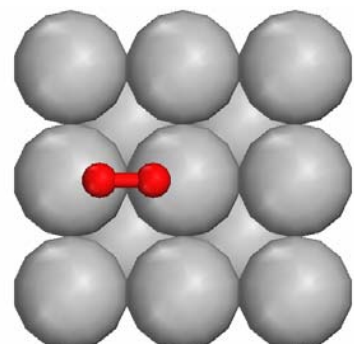

NSP  SP

(b) Hor1 approach.

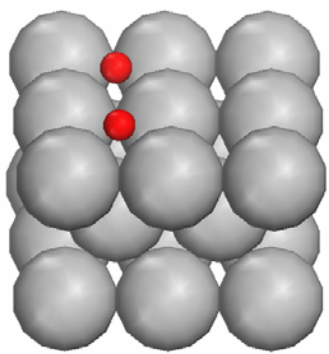
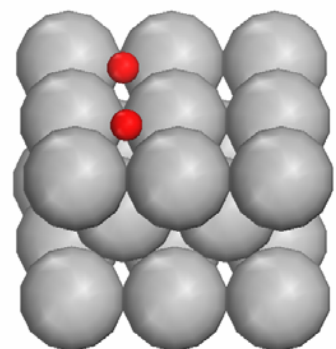

NSP  SP

(c) Hor2 approach.

Figure 2. Optimized configuration of $O_2$ adsorption on bridge site at different approaches. NSP and SP refer to non-spin and spin polarization, respectively.

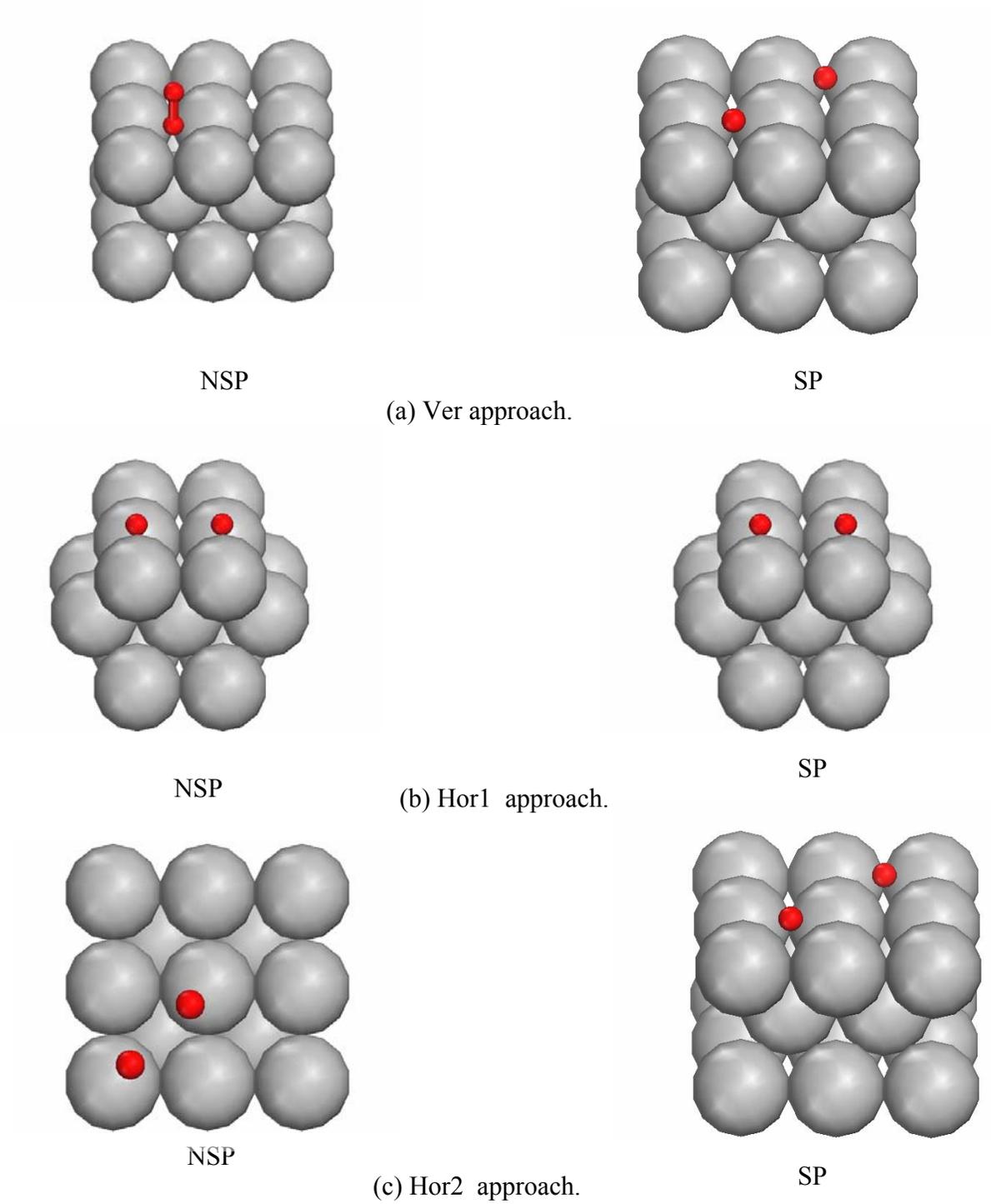

Figure 3. Optimized configuration of $O_2$ adsorption on center site at different approaches. NSP and SP refer to non-spin and spin polarization, respectively.

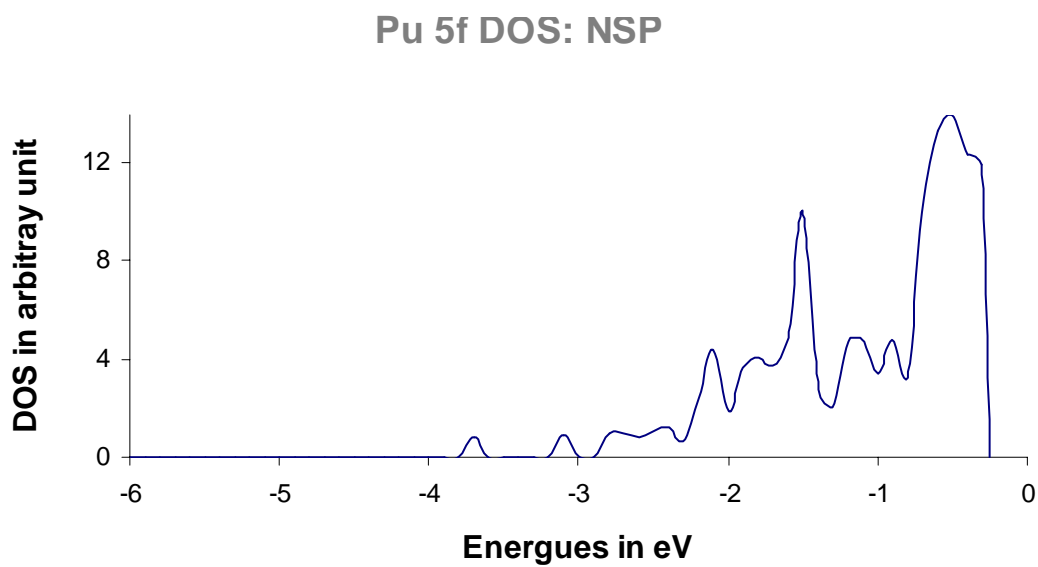

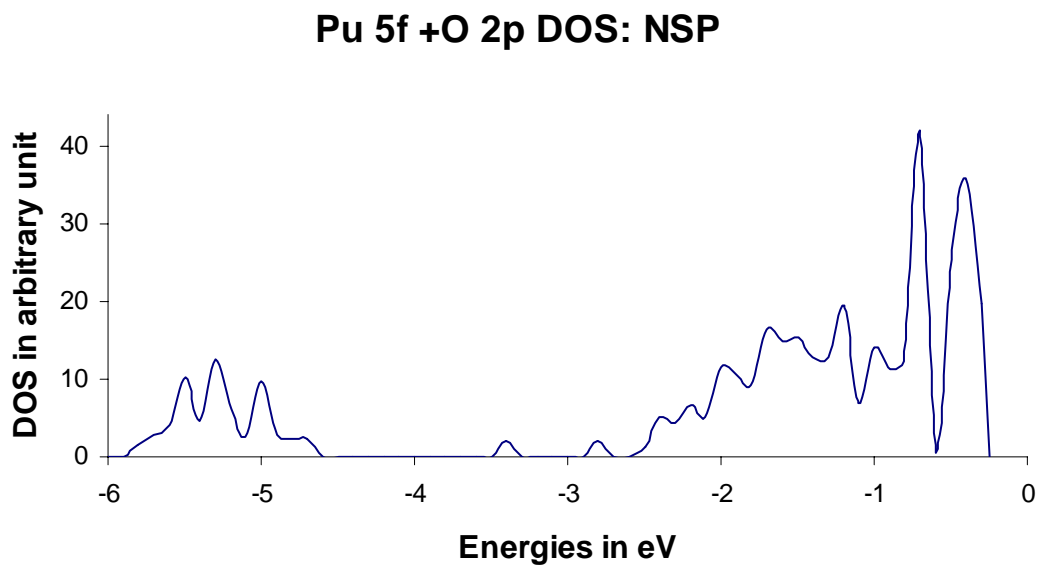

Figure 4: Density of states of 5f orbitals of plutonium (100) surface without and with oxygen at non-spin-polarized (NSP) level for the most favorable site, at *Hor2* approach on bridge site. Fermi energy is normalized to zero.

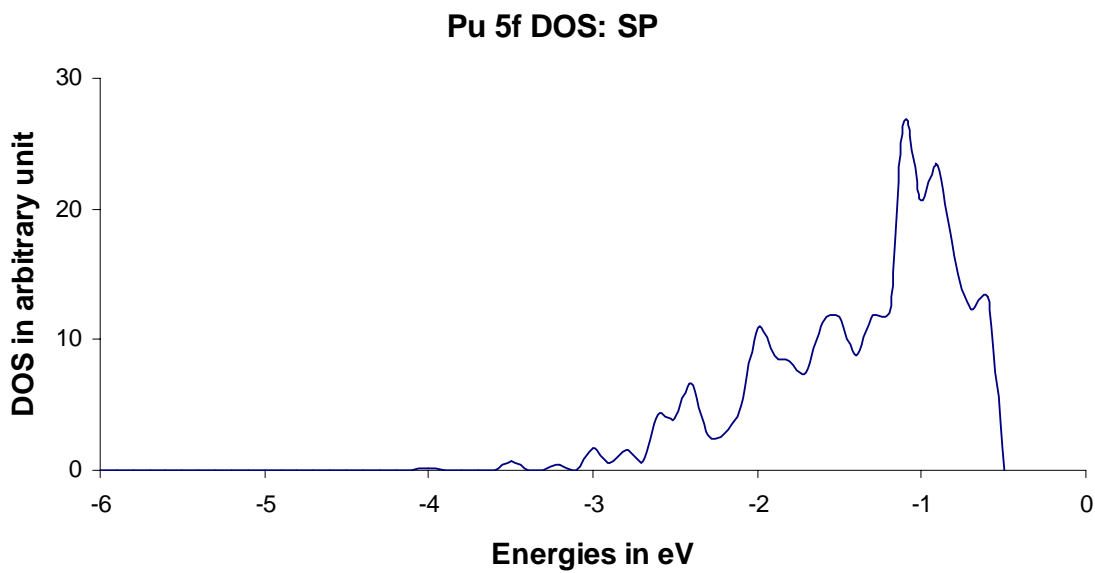

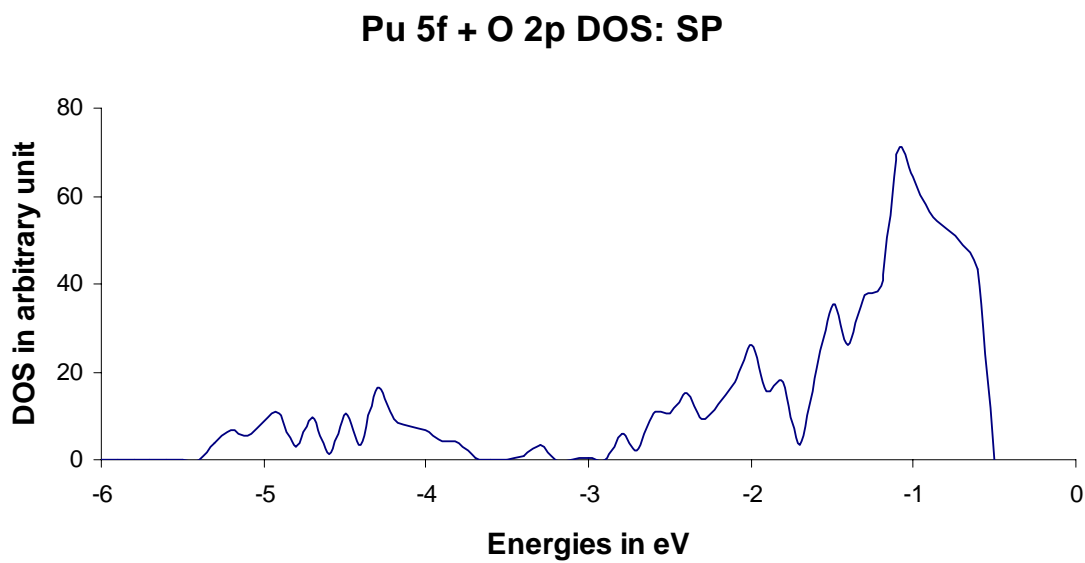

Figure 5. Density of states of 5f orbitals of plutonium (100) surface with and without oxygen at spin polarized (SP) level for the most favorable site, at *Hor2* approach on center site. Fermi energy is normalized to zero.

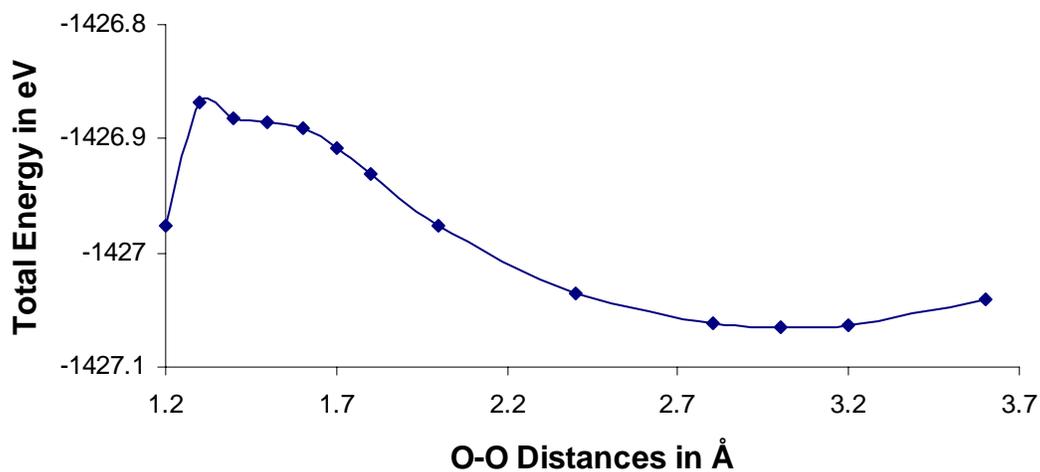

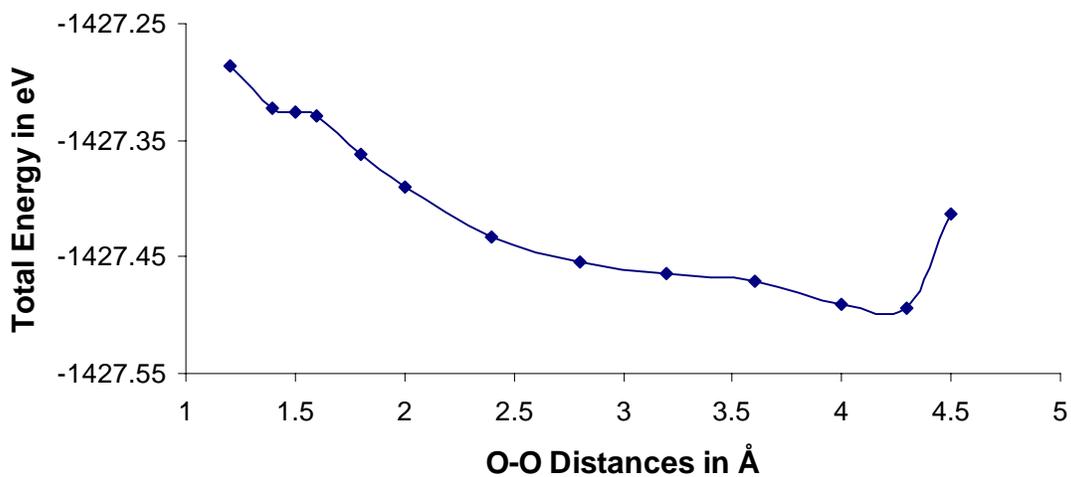

Figure 6. Reaction barriers for the most favorable non-spin-polarized (NSP) and spin-polarized (SP) sites. O-O distances, starting from experimental bond length of 1.207 Å, were used as reaction coordinates.